\documentclass[
    ,final            
  ]
  {aipproc}

\layoutstyle{6x9}

\begin{document}

\title{Performance of the ATLAS Detector on First Single Beam and Cosmic Ray Data}


\author{M. Aleksa on behalf of the ATLAS collaboration}{
  address={CERN, PH Department, Geneva 23, 1211-Geneva, Switzerland}
}

\begin{abstract}
We report on performance studies of the ATLAS detector obtained
with first single LHC (Large Hadron Collider) beam data in September 2008, and large samples
of cosmic ray events collected in the fall of 2008. In particular,
the performance of the calorimeter, crucial for jet and missing
transverse energy measurements, is studied. It is shown that the 
ATLAS experiment is ready to record the first LHC collisions.
\end{abstract}

\maketitle


\section{The ATLAS Experiment}

ATLAS~\cite{ATLASpaper} is a general purpose detector built to study high-energy proton-proton 
collisions at the LHC. 
Commissioning of the detector has 
started in 2005 in parallel to its installation. Large samples of cosmic muons have been recorded, 
which are used to understand and improve the performance of the detector, in particular for 
detector alignment and for first calibrations. The data are very useful to test cable mappings 
and the channel and trigger timing, to determine dead and noisy channels, and to verify the stability of the hardware 
during operation. Data were also recorded in September 2008 with first 
single beams circulating in the LHC.
%
%
The ATLAS experiment consists of an inner tracking detector, a calorimeter system, and an outer muon system. In the following sections there is a small overview of the readiness of the different detector parts. A more detailed description can be found in~\cite{thilo-ATLAS}.
%
%

The Inner Detector (ID) consists of the Pixel detector, the Semiconductor Tracker (SCT) and the Transition Radiation Tracker (TRT), which are operated inside a 2\,T solenoidal magnetic field parallel to the beam axis. It has a coverage in pseudo-rapidity, $\eta$, of $|\eta| < 2.5$ (TRT $|\eta| < 2$) and a momentum resolution of $p_T\cdot\sigma(q/p_T)=0.04\,\%\cdot (p_T/\mathrm{GeV})\oplus 1.5\,\%$ at $\eta=0$.
The Pixel Detector consists of three layers in the barrel and end-cap regions, with 80 million pixels of size $50\,\mu\mathrm{m}\times 400\,\mu\mathrm{m}$. More than 98\,\% of the modules are operational.
The SCT consists of four double layers of $80\,\mu\mathrm{m}$ thin silicon strips in the barrel region and nine in each end-cap. It has a total of six million channels spread over 4088 modules. A fraction of 99.3\,\% of the modules are operational.
The single hit efficiency is larger than 99\,\%.
The TRT is a combined straw tube tracker and transition radiation detector, which allows electron-pion identification in the energy region between 500\,MeV and 150\,GeV. The straw tubes are 4\,mm in diameter and contain a 35\,$\mu\mathrm{m}$ thin anode wire. 
98.2\,\% of the channels are operational.
%
%

ATLAS includes two types of sampling calorimeters: The Liquid Argon (LAr) calorimeter and the Tile calorimeter. The LAr calorimeter comprises the electromagnetic calorimetry in the central, end-cap and forward regions, as well as the hadronic calorimetry in the end-cap and forward regions. The electromagnetic part consists of layers of lead and liquid argon in an accordion geometry, with three longitudinal compartments in the region of $|\eta| < 2.5$. A presampler detector assists in $|\eta| < 1.8$. The hadronic calorimeter in the end-cap regions and the forward calorimeter use copper and tungsten as absorber materials. The hadronic calorimeter in the central region, the Tile calorimeter, consists of iron absorbers with scintillating tiles as active material.
The expected noise subtracted electromagnetic energy resolution is $\sigma/E = 10\,\%/\sqrt{E/\mathrm{GeV}} \oplus 0.7\,\%$ and the expected hadronic (jet) energy resolution is $\sigma/E = 100\,\%/\sqrt{E/\mathrm{GeV}} \oplus 10\,\%$ for $3.1 < |\eta| < 4.9$. The expected hadronic (jet) energy resolution is $\sigma/E = 50\,\%/\sqrt{E/\mathrm{GeV}} \oplus 3\,\%$ for $|\eta| < 3.1$. 
In the LAr calorimeter about 0.02\,\% of channels were found to be dead, and an additional 1.2\,\% of channels were found to be dead but can be recovered at the next shutdown. 
The Tile calorimeter dead cells have been reduced to $0.5\,\%$ during last shutdown. 
%
%

The muon spectrometer consists of four different kinds of muon chambers covering the region of $|\eta| < 2.5$ and an air-core toroid magnet system of rigidity $1.5-5.5\,\mathrm{Tm}$ for $|\eta| < 1.4$ and $1-7.5\,\mathrm{Tm}$ for $|\eta| > 1.6$. The expected standalone momentum resolution is $\sigma/p_T < 10\,\%$ at 1\,TeV or $3.5\,\% - 4.0\,\%$ at 100\,GeV.
For triggering at Level-1, fast chambers are used with a time resolution below 10\,ns and two-dimensional readout with a space resolution of 5-10mm. The barrel region consists of resistive-plate chambers (RPC) and the endcap region of thin gap chambers (TGC), more than $97\,\%$ and $98\,\%$ of chambers are operational, respectively.
For precision tracking, precision chambers are used with a spatial resolution of $35-40\,\mu\mathrm{m}$. The barrel and end-cap regions are instrumented with 1088 chambers of monitored drift tubes (MDT). 
About 99.3\,\% of the MDT chambers are operational. The fraction of noisy channels with at least 5\,\% occupancy is less than 0.2\,\%. In the forward direction, 32 cathode-strip chambers (CSC) are used. A fraction of 98.4\,\% of the chambers are operational.

\section{Commissioning with Cosmic Rays}

From mid-Sept. 2008 until the end of Oct. 2008 over 200 million cosmic muon events were recorded with the ATLAS detector. The muon rate inside the cavern is about 4\,kHz inside the muon system fiducial volume (but only $\sim 15$\,Hz in the TRT barrel detector).
In the following a few selected results from cosmic muon analyses will be shown.
\begin{itemize}
\item {\bf Missing transverse energy studies with the calorimeter:} Randomly triggered events have been used to estimate the transverse missing energy $E_T^\mathrm{miss}$ defined as $(E_T^\mathrm{miss})^2=(\sum E\sin\theta\cos\varphi)^2+(\sum E\sin\theta\sin\varphi)^2$, where $\theta$ and $\varphi$ are the polar and azimuthal angle, respectively. Figure~\ref{fig:missingEt} shows $E_T^\mathrm{miss}$ with the sum calculated in two different ways (left plot). The squares, where only the energy inside ATLAS standard topological clusters is summed, show better noise suppression. The tails of the distribution above 8\,GeV can be explained by a specific presampler region that had large coherent noise during data taking. This noise problem has been fixed in the winter of 2009 after the data in Fig.~\ref{fig:missingEt} was recorded. The right plot shows the stability of the missing transverse energy, calculated with topological clusters, during the whole data taking period. 
\begin{figure}
\includegraphics[height=.17\textheight]{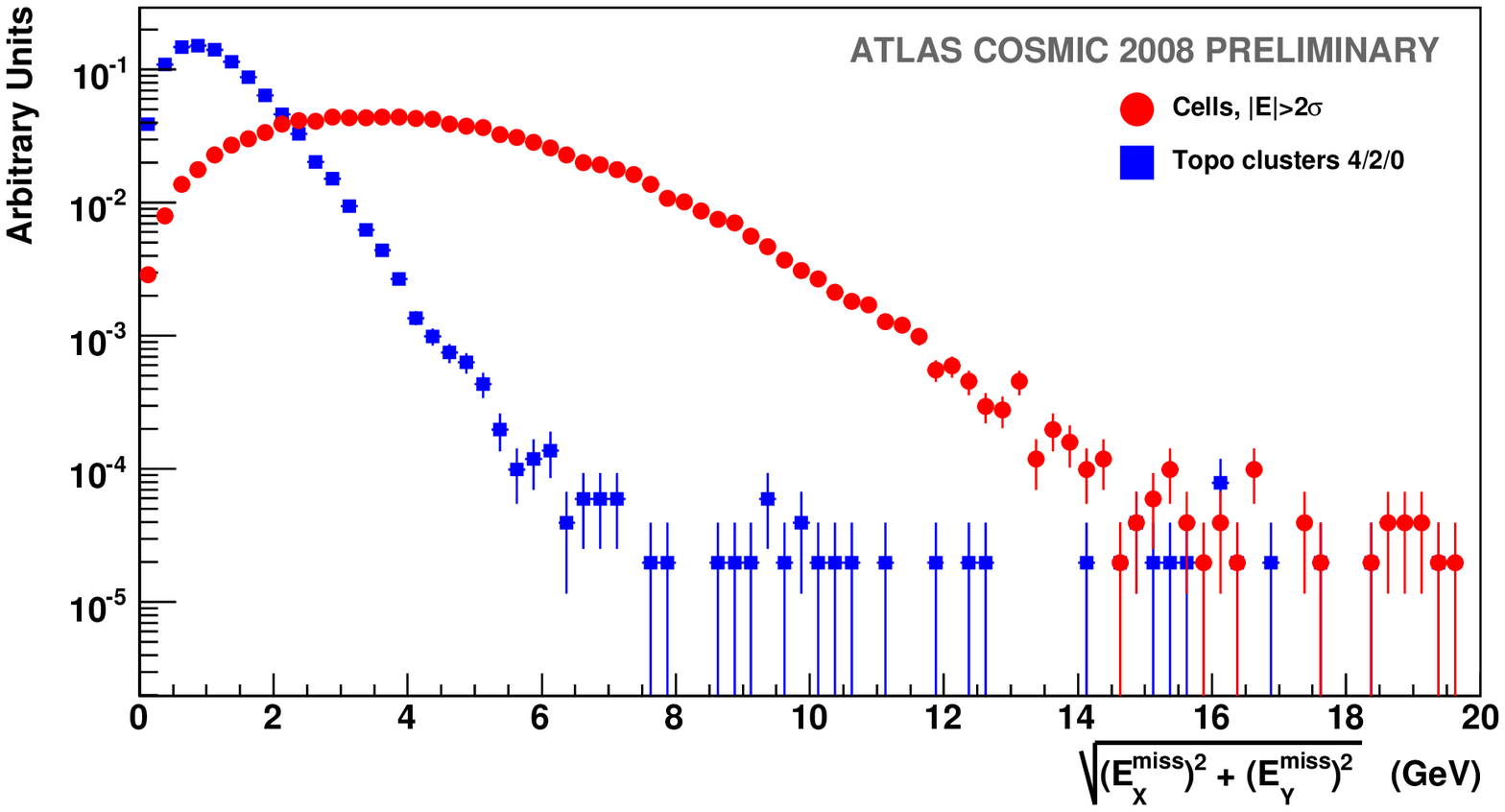}\quad\quad\includegraphics[height=.17\textheight]{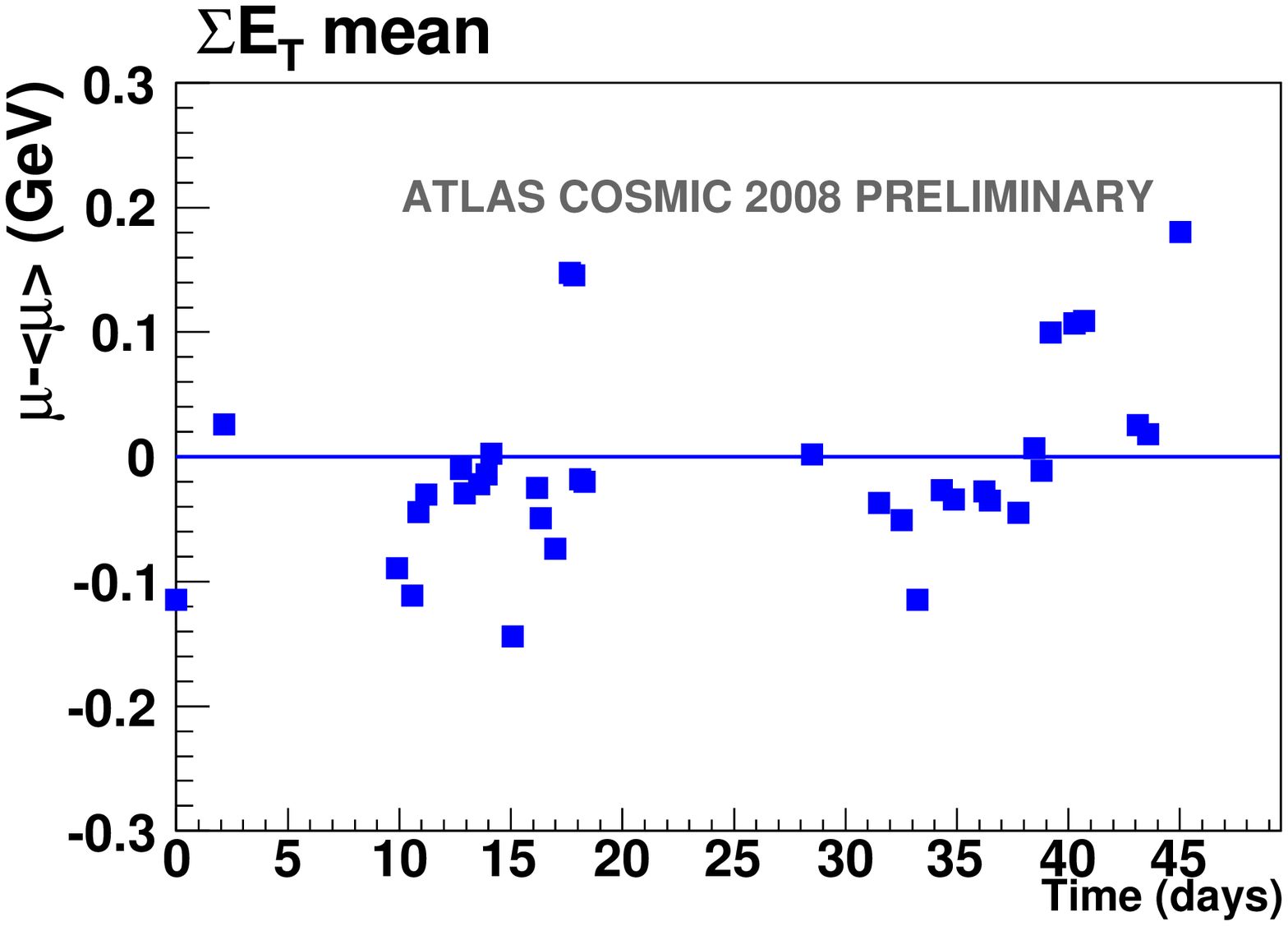}
  \caption{Left plot: Missing transverse energy $E_T^\mathrm{miss}$ in randomly triggered events. The red circles show the sum of all cells with an energy greater than two noise standard deviations. The blue squares correspond to standard topological clusters. Right plot: Stability of $E_T^\mathrm{miss}$ over the data taking period in Sept. and Oct. 2008.}\label{fig:missingEt}
\end{figure}
%
%
\item {\bf Combined tracking studies:} The track parameters of cosmic muon rays reconstructed in the muon system and in the ID have been compared. All the track parameters measured with the ID and the muon system agree well, and the width of the distributions are as expected, in good agreement with Monte Carlo detector simulations (MC). Figure~\ref{fig:comb-track} shows the correlation of the polar angle $\theta$ measured with the ID and the muon system (left plot). The right plot shows the difference of the muon momentum measured in the ID and in the bottom part of the muon system. The measured average difference of $\sim 3\,\mathrm{GeV}$ is due to the energy loss in the calorimeters. The mean and the width of the distribution are well described by the MC simulation.  
\begin{figure}
\includegraphics[height=.22\textheight]{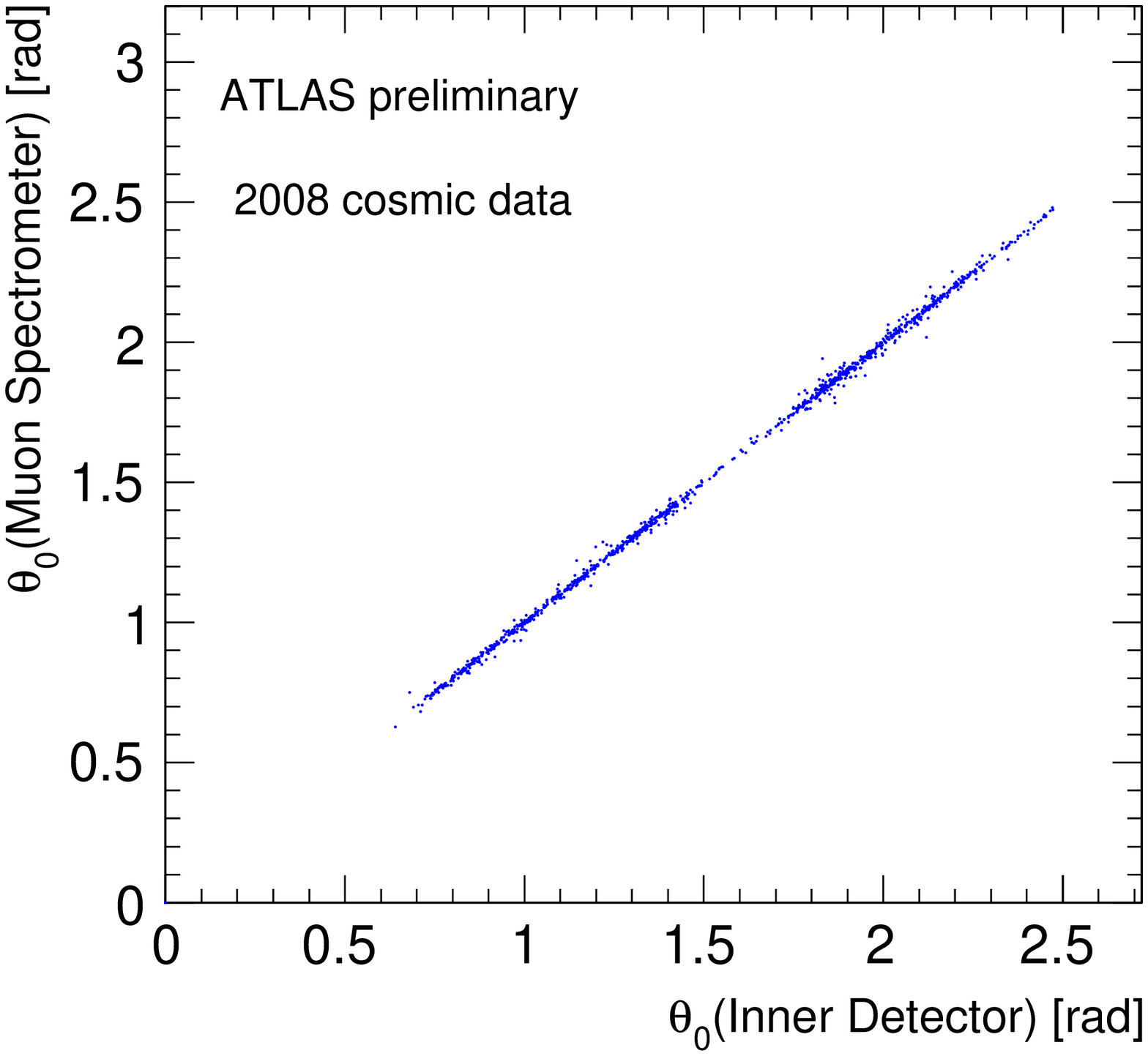}\quad\quad\quad\includegraphics[height=.22\textheight]{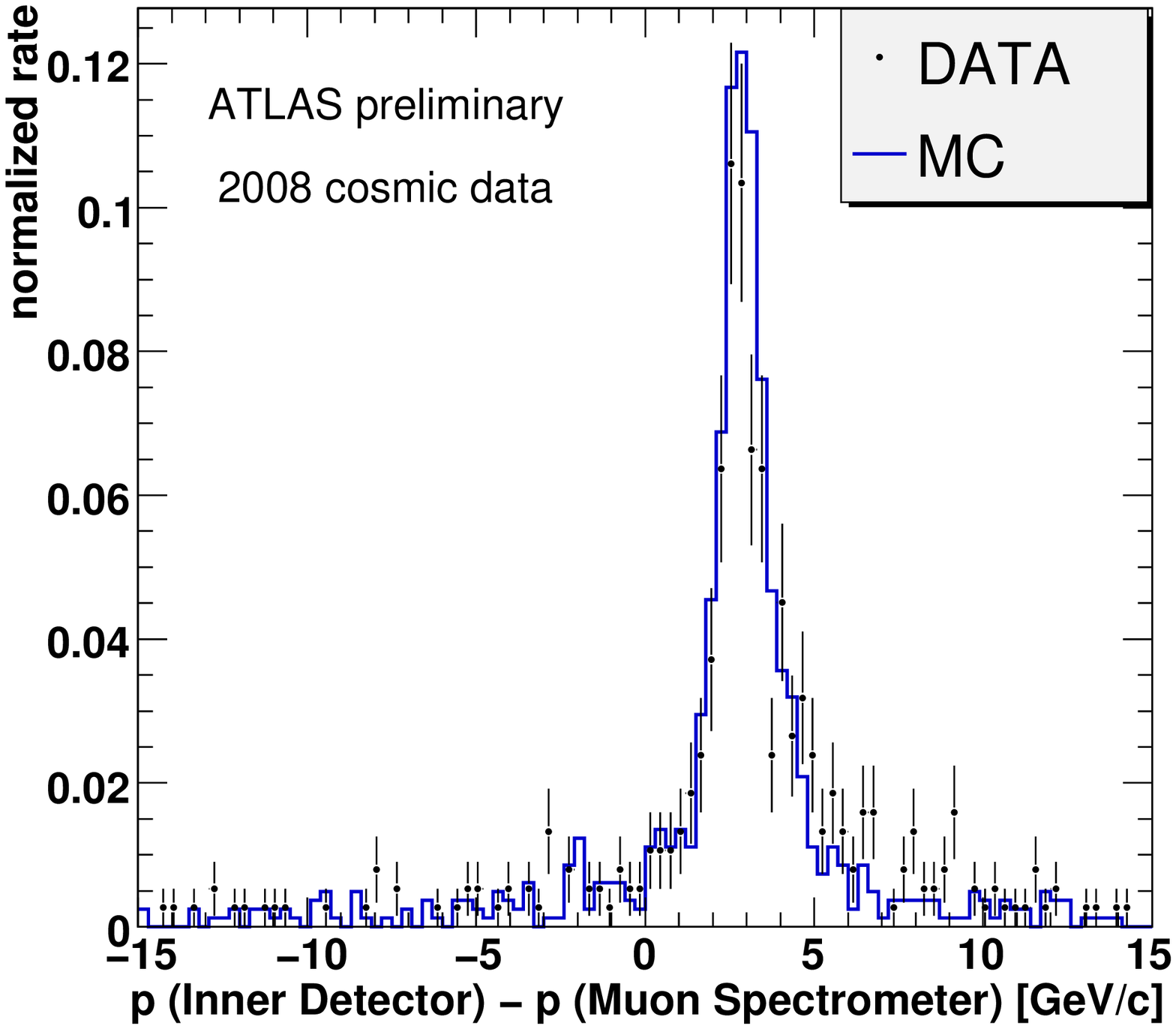}
  \caption{Left plot: Correlation of the polar angle $\theta$ measured with the ID and the bottom part of the muon system. Right plot: Difference of the muon momentum measured in the ID and in the muon system.}\label{fig:comb-track}
\end{figure}
\end{itemize}
\section{Commissioning with Beam}

On Sept. 10, 2008 LHC single beams went through ATLAS for the first time. About 100 beam ``splash'' events were produced by dumping the beam into a closed collimator 140\,m upstream of the ATLAS experiment and thus producing muon showers of hundreds of thousands of muons reaching the cavern. ATLAS\footnote{Almost all ATLAS sub-detectors participated. However, due to detector safety the SCT barrel and the Pixel detector were switched off. The SCT endcap, the forward calorimeter and some parts of the muon system were operated with reduced high voltage settings.} recorded most of these events, which produced thousands of track segments in the tracking detectors and remarkable energy deposits of up to 1000\,TeV in the calorimeters.
\begin{itemize}
%
\item {\bf Timing with beam splash events:} Assuming that all the muons produced in a beam splash propagate along the beam direction and arrive in the cavern at the same time, within a few nanoseconds, beam splash events are ideal to check the timing of the trigger system and the read-out of the different sub-detectors. After only one day the trigger timing was aligned to be within one bunch crossing (i.e. 25\,ns). After detailed analysis of the data, the TRT detector was timed in at the 1\,ns level. The left plot in Fig.~\ref{fig:timing-splash} shows the $\varphi$-average of the time response of the different longitudinal layers (samples) of the Tile calorimeter. It can be seen that after a time-of-flight correction the different partitions have time dispersions of less than 2\,ns. The results of this study have been used to align the different partitions with respect to each other. The right plot in Fig.~\ref{fig:timing-splash} shows the result of a timing study with the LAr barrel calorimeter. The left crosses in each bin (slot) denote the expected times of the signal peaks for the different front-end electronics boards\footnote{Each board covers a specific longitudinal compartment and region in pseudo-rapidity $\eta$.}
calculated from the calibration system and the read-out path. They are compared to the measured signal times during beam splash events (time-of-flight corrected, $\varphi$-average). The agreement is at the 2\,ns level apart from the presampler; the discrepancy has since been explained. These corrections and an additional correction for the time of flight from the interaction point have since been applied in the read-out electronics.
\begin{figure}
\includegraphics[height=.16\textheight]{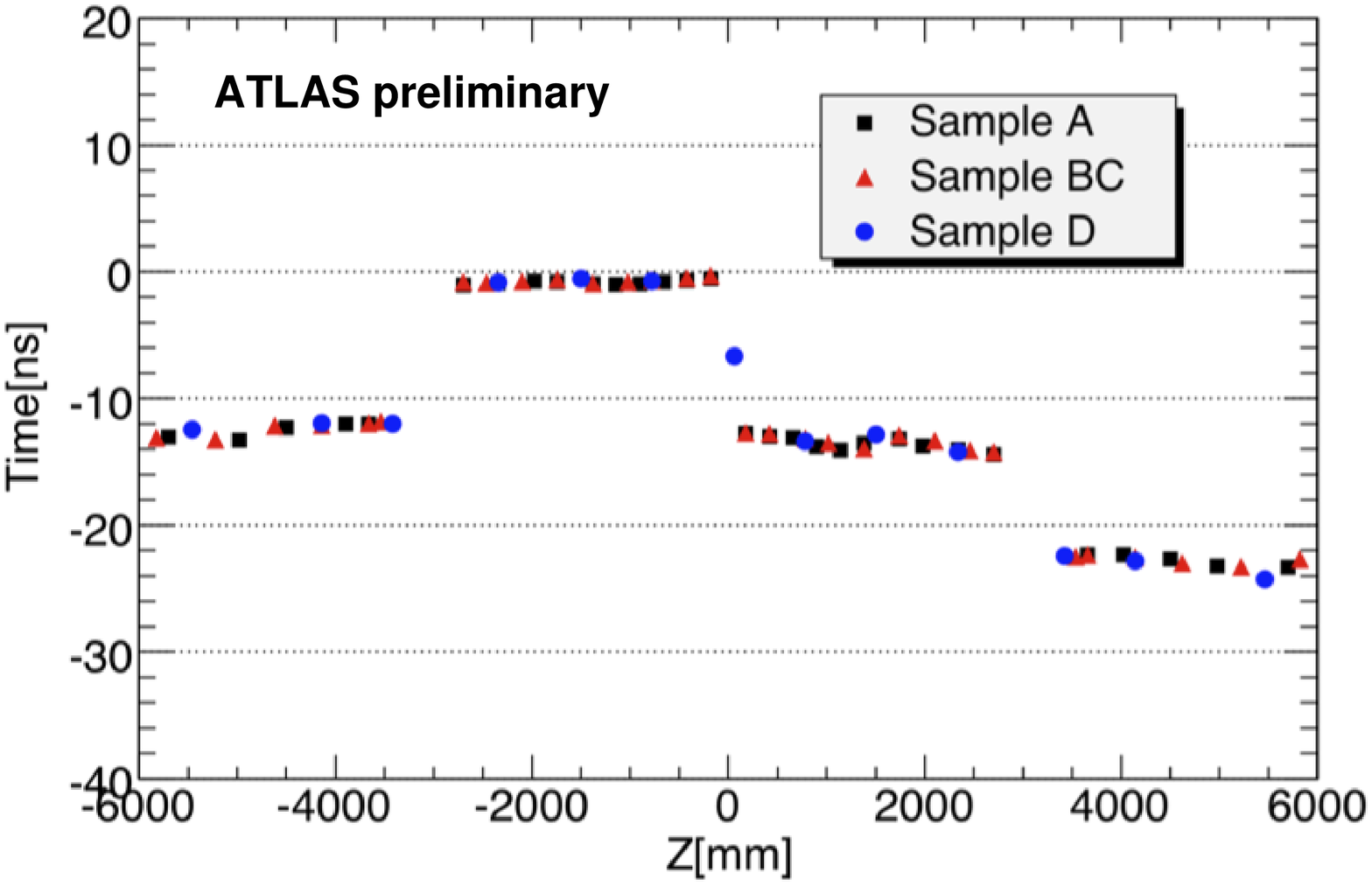}\quad\quad\includegraphics[height=.16\textheight]{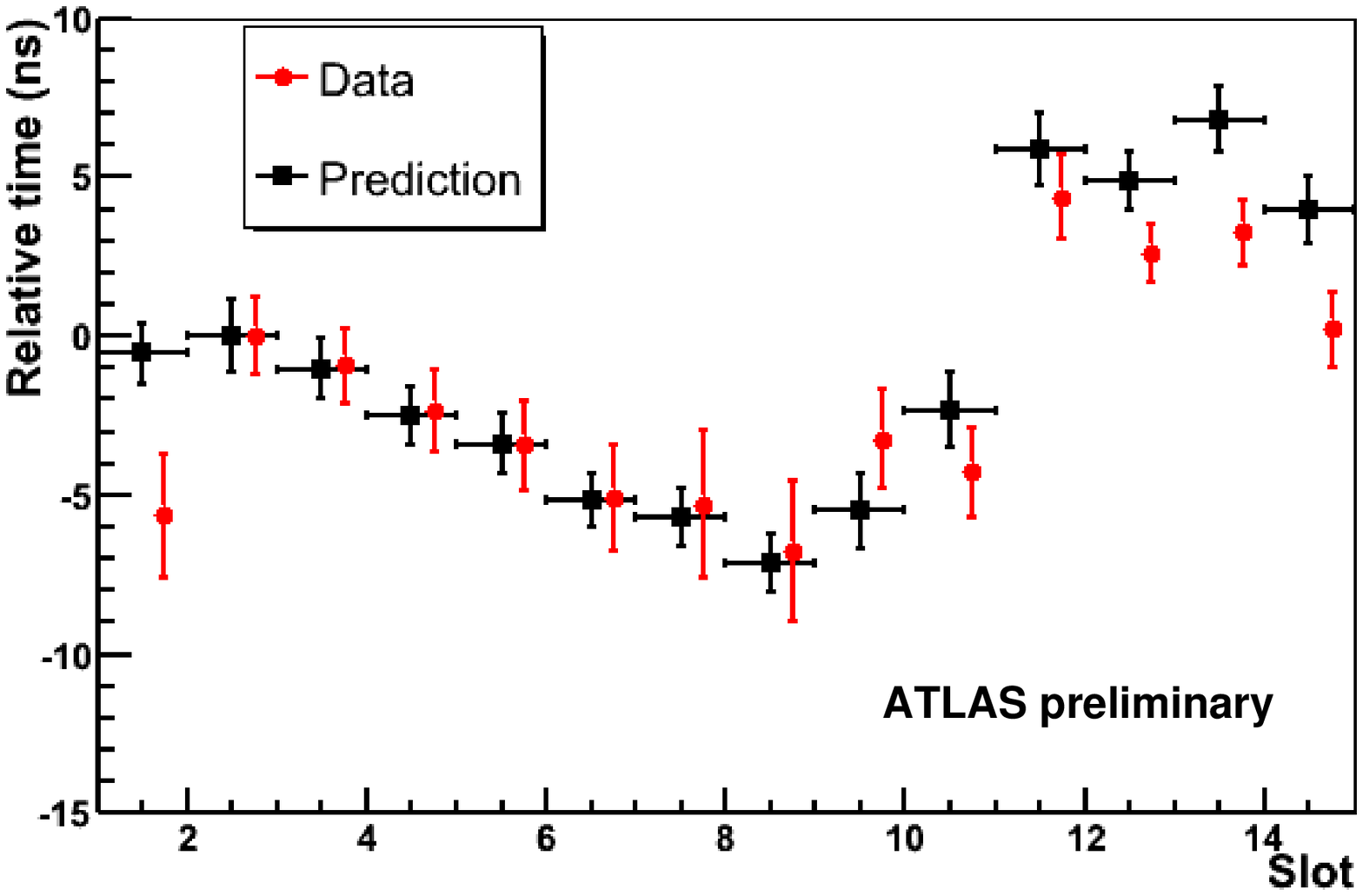}
  \caption{Left plot: Time response of the different longitudinal layers (samples) of the Tile calorimeter. Right plot: Comparison between the expected and measured signal timing for the different LAr calorimeter barrel front-end-electronics boards.}\label{fig:timing-splash}
\end{figure}
\end{itemize}






\bibliographystyle{aipproc}   

\bibliography{References}

\end{document}